# Flexible and Cost-Effective Spherical to Cartesian Coordinate Conversion Using 3-D CORDIC Algorithm on FPGA

**Nadia M. Salem[1], Sami I. Serhan[2], Khawla M. Al-Tarawneh[3], Ra'fat Al-Msie'deen[4]**

**Abstract**: In computer science, transforming spherical coordinates into Cartesian coordinates is an important mathematical operation. The CORDIC (Coordinate Rotation Digital Computer) iterative algorithm can perform this operation, as well as trigonometric functions and vector rotations, using only simple arithmetic operations like addition, subtraction, and bit-shifting. This research paper presents hardware architecture for a 3-D CORDIC processor using Quartus II 7.1 ALTERA software, which enables easy modifications and design changes due to its regularity and simplicity. The proposed 3-D CORDIC model is evaluated by comparing the calculated results with the simulated results to determine its accuracy. The results were satisfaction and the proposed model could be suitable for numerous real-time applications.

## 1. Introduction

The iterative algorithm called: CORDIC Coordinate Rotation Digital Computer calculates the values of trigonometric functions like sine, cosine, square root, logarithm, magnitude, phase and more functions. [1].

The CORDIC algorithm is a method that rotates a vector repeatedly to approximate mathematical functions. It accomplishes this by decomposing the desired function into a series of minute rotations and shifts. The vector rotates by a predetermined degree on each iteration, and its coordinates are updated accordingly. Multiple iterations of this procedure are carried out until the desired precision is achieved. Because of CORDIC's efficiency and simplicity, hardware and embedded systems can benefit from its use [2].

In computer science, transforming spherical coordinates to Cartesian coordinates is a crucial mathematical operation with applications in computer graphics [14], game development, robotics, and signal processing [5], Barcode Identification [15], fingerprints recognition systems [10]. The 3-D CORDIC algorithm is a well-known method for this conversion, rotating the spherical coordinate system and approximating the conversion with high precision through using straightforward shift and adds operations. Due to the CORDIC algorithm's regularity and simplicity for generating high throughput and low latency, implementing it using an FPGA provides a flexible and affordable development environment, allowing for easy design revisions [11]. The proposed 3-D CORDIC processor that converts from spherical to Cartesian coordinates was implemented using the (Very High-Speed Integrated Circuit Hardware Description Language) VHDL. First of all build and installation of memory entity (angle) to hold the arctangent ,then Build the 2-D CORDIC entity with angle LUT memory entity, then compile and simulate the entity functions using Quartus II 7.1 tool of ALTERA. The last step construct 3-D CORDIC Processor using two entities of 2-D CORDIC and compile and simulate the entity functions using Quartus II 7.1 tool of ALTERA.

VHDL stands for Very High-Speed Integrated Circuit Hardware Description Language. It is a programming language used to describe and simulate digital circuits and systems. VHDL allows for the design and modeling of hardware components, such as processors, controllers, and other digital systems, using a concise and structured syntax. It is commonly used in the field of digital design for FPGA and ASIC implementations [21].

The remaining paper is organized as follows:

Section 2 presents the prior work for various CORDIC architecture types. In Section 3 experimental work includes the theory of CORDIC algorithm and spherical coordinates and the proposed architecture for rotation mode derived from the algorithm for 2-D CORDIC and 3-D CORDIC processor are presented. In section 4 the results and discussion about the simulation of the proposed model on VHDL are reported. Finally, section 5 have the conclusions.

## 2. Related works

The Cordic algorithm has a rich history dating to the 1950s when it was first introduced by Jack Volder. It was initially developed as a method to perform trigonometric calculations using digital hardware. Over the years, the algorithm has been refined and expanded, finding widespread application in various fields such as navigation, signal processing, and graphics due to its efficiency and versatility [2].

[1] *King Abdullah II School for Information Technology, Jordan's University, nad9220478@ju.edu.jo*
[2] *King Abdullah II School for Information Technology, Jordan's University, samiserh@ju.edu.jo*
[3] *King Abdullah II School for Information Technology, Jordan's University, Kol9220471@ju.edu.jo*
[4] *Faculty of Information Technology, Mutah University, rafatalmsiedeen@mutah.edu.jo*

Jack Volder [1] presented the Coordinate Rotation Digital Computer (CORDIC) which is designed for real-time airborne computation. It utilizes a distinct computing technique that is particularly suited for resolving the trigonometric relationships required for plane coordinate rotation and rectangular to polar coordinate conversion. Volder compares the limitations of existing methods, explains how CORDIC overcomes these, and covers various forms of the algorithm. Volder highlighted the algorithm's accuracy and efficiency and its applications in signal processing, control systems, and computer graphics. Compared the limitations of existing methods, explained how CORDIC overcomes these, and covers various forms of the algorithm.

Jack Volder [2] traced the history and development of the CORDIC algorithm, a technique used for computing basic functions such as sine, cosine, and logarithms. The article provides a detailed explanation of the algorithm's derivation and the motivation behind its creation which was the need of B-58 aircraft's analogy navigation computer to be replaced with a high-accuracy and high-performance digital computer. The author also discussed the advantages and disadvantages of the CORDIC algorithm compared to other methods, and highlighted its simplicity and efficiency in a wide range of applications. Overall, the article provides a comprehensive overview of the CORDIC algorithm and its importance in modern computing.

Kumar [6] discussed the implementation of serial and parallel architectures for several mathematical functions (Sine, Cosine, Exponential, Inverse Exponential, Logarithm and Rectangular to polar) on the Cyclone IV E FPGA, with a focus on comparing the area, delay, and power consumption metrics of each architecture. The results indicate that the serial architecture is more area-efficient, while the parallel architecture requires more area. However, the parallel architecture outperforms the serial design in terms of speed. This trade-off between latency and accuracy can be beneficial for various real-time applications.

Mazenc et al. [7] presented an extension of the Coordinate Rotation Digital Computer algorithm, enabling the computation of several functions, including $\cos^{-1}$, $\sin^{-1}$, $\cosh^{-1}$ and $\sinh^{-1}$.

Walther [9] presented Coordinate Rotation Digital Computer algorithm for computing elementary functions such as multiplication, division, sin, cos, tan, arctan, sinh, cosh, tanh, arctanh, exp and square-root. The author also describes a hardware floating point processor built using the algorithm at Hewlett-Packard Laboratories, complete with a block diagram, microprogram control details, and actual performance metrics.

Lakshmi et al. [13] A Survey implied that CORDIC algorithm has two main advantages reduced latency and improved throughput.

Sergiyenko et al. [16] proposed a new algorithm for calculating sine and cosine functions using three stages of rotations based on a modified hybrid approach. These stages involve using a ROM table, a network of CORDIC micro rotations. The error in calculations for small angles is within acceptable limits.

H. Nair and A. Chalil [17] implemented a 32-bit floating-point serial and parallel CORDIC architecture on an FPGA using different adders to create a more area and speed efficient CORDIC architecture. The proposed architecture with Ladner Fischer adder has better area utilization and less delay compared to other adders in both serial and parallel CORDIC architectures

Paz and Garrido [18] proposed that new CORDIC-based algorithm does not require complex iterations or actual multiplication to compute functions accurately, unlike previous approaches. The proposed algorithm has been implemented in hardware and shown to have a better balance between space usage and accuracy compared to other CORDIC-based approaches. The authors make a comparison between different pipelined architectures used for the computation of arcsine using CORDIC.

Wang et al. [19] proposed a design for a high-accuracy and energy-efficient Izhikevich neuron based on Fast-Convergence Coordinate Rotation Digital Computer. The design includes an error propagation model for systematic error analysis and effective error reduction, along with two methods for reducing errors in the design. By using FC-CORDIC for square calculation, redundant CORDIC iterations are eliminated, improving accuracy and energy efficiency.

K. T. Chen et al. [20] mentioned that many functions can be computed by CORDIC algorithm as shown in the table 1 below :( **\*this is our proposed model idea)**

**Table 1:** CORDIC algorithm calculated functions [20].

| Coordinate system(m) | Rotation Mode $Z_n=0$ | Vectoring Mode $X_n=0$ or $Y_n=0$ |
|---|---|---|
| **Circular\*** **m=1** | **sin(x)** **cos(x)** **tan(x)** | arcsin(x) arccos(x) arctan(x) |

| Linear m=0 | f(x,y)=xy | f(x,y)=x/y |
|---|---|---|
| Hyperbolic m=-1 | sinh(x) cosh(x) tanh(x) $e^x$ | arcsinh(x) arccosh(x) arctanh(x) ln(x) $x^{0.5}$ |

Table2 presents a summary of some articles about CORDIC algorithm.

**Table 2: Some CORDIC algorithm articles.**

| Researcher | Article's Main topic in CORDIC |
|---|---|
| Kumar[6] 2019 | Compared serial and parallel architectures for mathematical functions and found that the parallel architecture offers better speed but requires more area. |
| Mazenc et al.[7] 1993 | Extended the algorithm to compute additional functions such as inverse trigonometric and hyperbolic functions |
| Walther [9] 1971 | Presented the CORDIC algorithm for elementary functions and described a hardware floating-point processor implementation |
| Lakshmi et al. [13] 2010 | Highlighted the advantages of the CORDIC algorithm, including reduced latency and improved throughput |
| Sergiyenko et al. [16] 2021 | Introduced a modified hybrid approach for calculating sine and cosine functions with reduced hardware volume and calculation delay. |
| Nair and Chalil [17] 2022 | Implemented a more efficient CORDIC architecture using different adders |
| Paz and Garrido [18] 2023 | Proposed a new CORDIC-based algorithm that eliminates the need for complex iterations or actual multiplication for accurate function computation. |
| Wang et al.[19] 2022 | Designed a high-accuracy and energy-efficient neuron based on the Fast-Convergence Coordinate Rotation Digital Computer, which eliminates redundant iterations and improves accuracy and energy efficiency. |

It can be said that the CORDIC algorithm has a significant impact on modern computing, providing efficient and accurate computation for various functions in diverse applications. Researchers continue to explore and enhance the algorithm for improved performance and application-specific optimizations. Finally, it's obvious that CORDIC reduced latency and improved throughput [19].

## 3. Experimental Procedure

### 3.1 Theory

#### 3.1.1 Spherical Coordinates System [8]

The spherical coordinate system is a way to describe the location of a point in 3D space using three values: how far the point is from a fixed starting point (radial distance), how high or low the point is compared to a fixed up-down direction (inclination or elevation angle), and the direction of the point's projection onto a flat plane that passes through the starting point and is perpendicular to the up-down direction (azimuth or reference angle). The inclination angle can also be replaced by an elevation angle measured from the flat reference plane. As seen in figure 1.

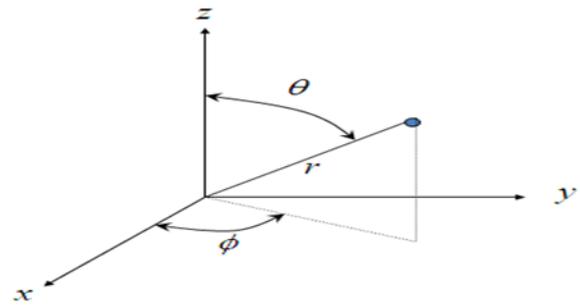

**Fig 1:** Spherical Coordinates [8]

A set of three values (r, θ, φ) in the spherical coordinate system tells us the exact location of a point in 3D space.

Spherical coordinate takes its importance because they are related to longitude (θ) and latitude (φ) which widely used in navigation systems. Based on this fact we can transfer any point on the earth (R, θ, φ) to its equivalent Cartesian coordinates (X, Y, Z).

To plot this point using its spherical coordinates, we follow these steps:

1- Move r units away from the starting point (origin) in the direction of the zenith (upward).

2- Rotate by θ degrees around the origin in the direction of the azimuth reference.

3- Rotate by φ degrees around the zenith (upward) in the correct direction.

These steps help us determine the precise location of the point in 3D space based on its spherical coordinates (r, θ, φ), where θ represents inclination or elevation angle. To convert spherical coordinates (r, θ, φ) to Cartesian coordinates (x, y, z), the following equations can be used [8]:

$$x = r \sin\theta \cos\Phi \quad (1)$$

$$y = r \sin\theta \sin\Phi \quad (2)$$

$$Z = r \cos\theta \quad (3)$$

### 3.1.2 CORDIC Algorithm

Volder's algorithm is based on general equations for rotating a vector in 2D space. Given a vector V with coordinates (x, y) that needs to be rotated by an angle Ø, we can get a new vector V' with updated coordinates (x', y') using the following method, which is derived from the equations for vector rotation [4]:

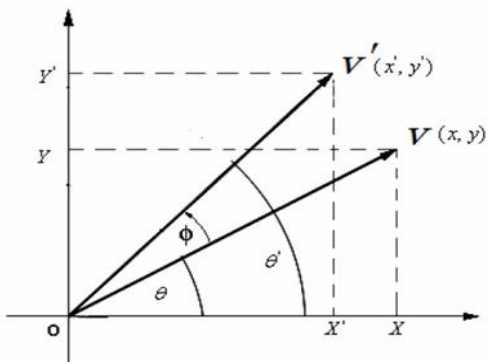

**Fig 2:** Vector V is rotated by angle Ø. [4]

$$x' = \cos(\Phi).[x - y.\tan(\Phi)] \quad (4)$$

$$y' = \cos(\Phi).[y + x.\tan(\Phi)] \quad (5)$$

$$(\theta)_{16} = \left\{ (\theta)_{10} * \left( \frac{16^4}{720} \right)_{10} \right\}$$

Rewrite in terms of αi: (0 ≤ i ≤ n) where **i** is iteration number [4]:

$$x_{i+1} = \cos(\alpha_i).[x_i - y_i.d_i.\tan(\alpha_i)] \quad (6)$$

$$y_{i+1} = \cos(\alpha_i).[y_i + x_i.d_i.\tan(\alpha_i)] \quad (7)$$

$$x_{i+1} = K_i.[x_i - y_i.d_i.2^{-i}] \quad (8)$$

$$y_{i+1} = K_i.[y_i + x_i.d_i.2^{-i}] \quad (9)$$

$$K_i = \cos(\theta_i) = \cos(\tan^{-1}(2^{-i})) = 1/\sqrt{1+2^{-2i}}$$
$$d_i = \pm 1$$

The product of the $K_i's$ approaches $0.607252935...$ as the number of iterations goes to infinity. For 11 iterations

K =cos(45.00000°) * cos(26.56505°) * cos(14.03624°) * cos(7.12502°) * cos(3.57633°) * cos(1.78991°) * cos(0.89517°) * cos(0.44761°) * cos(0.22380°) * cos(0.11190°) * cos(0.05590°) = $0.607252935...$

The exact gain depends on the number of iterations, according equation below:

$$K = \prod_{i=0}^{N-1} \sqrt{1/(1+2^{-2i})} \quad (10) \quad [4]$$

According to Volder's algorithm the CORDIC arithmetic unit is simple as shown in figure 3 below:

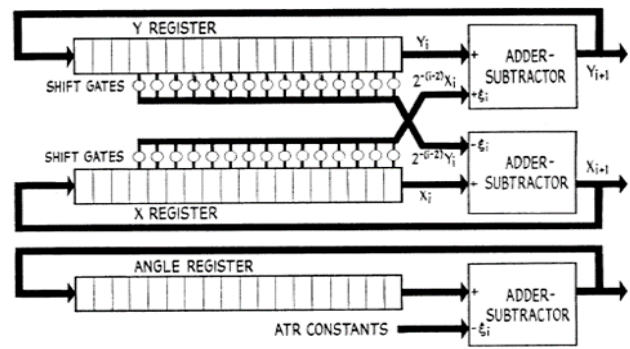

**Fig 3:** CORDIC Hardware. [1]

### 3.2 Implementation

#### 3.2.1 Angle Conversion Formula

This paper assumes that the conversion formula as follows:

Step size ◻ 1 degree = (2^16)/720 = 91.02 decimal = (5B)hex.

Assume that Θ =30 then t its equivalent in hexadecimal.

(30) hex. = {[(2^16)/720]*30}decimal=(2730)decimal=(0AAA)hex.

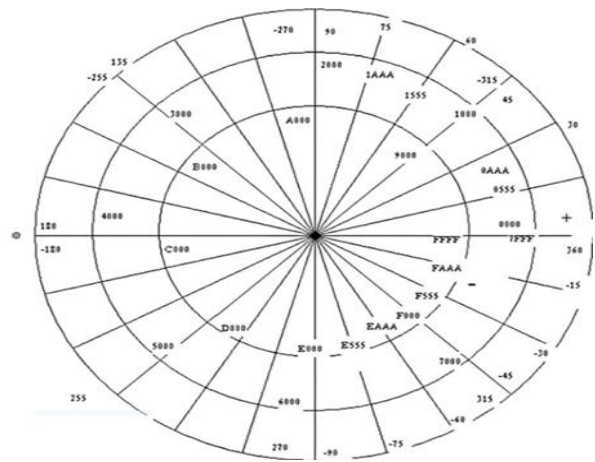

**Fig 4:** Decimal to Hex-decimal conversion angles in 2's complement representation.

Angles in CORDIC algorithm assume that θi such that, tan(θi)=2^-i

**Table 3:** Look up Table for angles used in CORDIC.

| I | tan(θ$_i$)=2$^{-i}$ | $\theta_i$ | Hexadecimal Value |
|---|---|---|---|
| 0 | 1 | 45.00000 ° | 1000 |
| 1 | 0.5 | 26.56505 ° | 9720 |
| 2 | 0.25 | 14.03624 ° | 4FD9 |
| 3 | 0.125 | 7.12502 ° | 2888 |
| 4 | 0.0625 | 3.57633 ° | 1458 |
| 5 | 0.03125 | 1.78991 ° | A2EB |
| 6 | 0.015625 | 0.89517 ° | 517B |
| 7 | 0.0078125 | 0.44761 ° | 28BE |
| 8 | 0.00396625 | 0.2238° | 145F |
| 9 | 0.001953125 | 0.1119° | A2F9 |
| 10 | 0.0009765625 | 0.0559° | 517C |
| 11 | 0.00048828125 | 0.0279° | 28BE |
| 12 | 0.000244140625 | 0.0139° | 145F |
| 13 | 0.0001220703125 | 0.0069° | A2F9 |
| 14 | 0.000061035156 | 0.0034° | 517C |
| 15 | 0.000030517578 | 0.0017° | 28BE |

In this paper first building and installation of lookup tables as shown below to hold the arctangent base angles where:

$$angles = \tan^{-1} 2^{-i} \quad where \quad 0 \leq i \leq 15$$

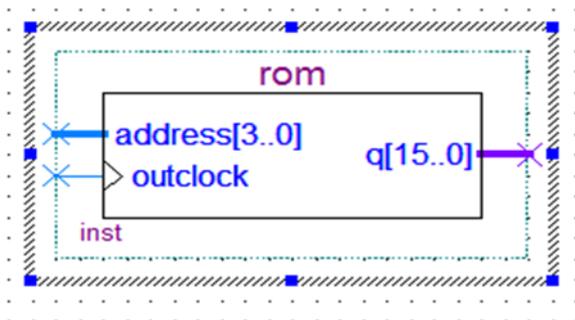

**Fig 5:** Angle LUT Input/output

### 3.2.2 Two-D CORDIC

In this paper 3-D CORDIC processor contains two cascaded 2-D CRDIC Processor that transforms from polar to Cartesian coordinates which is defined by the following equation [5]:

$$x = r\cos\theta \quad (11)$$

$$y = r\sin\theta \quad (12)$$

As pointed out above, the transformation is accomplished by selecting the rotation mode.

Let **X0= polar magnitude**, **z0= polar phase, y0=0** and using proposed 2-D CORDIC entity see Fig 6 below, the output will be rcosθ and rsinθ the result represents the polar input transformed to Cartesian space.

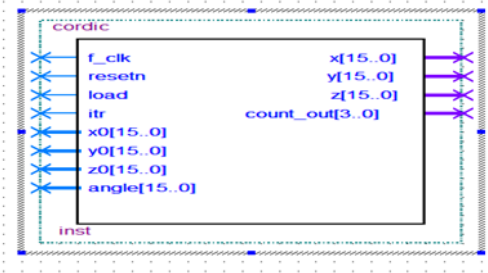

**Fig 6:** 2-D CORDIC Entity using VHDL

In this paper the proposed model used Rotation Mode of CORDIC algorithm as shown in the following table:

**Table 4: Some of CORDIC Rotation Mode Applications.**

| Function | Input | Output |
|---|---|---|
| cosϴ | X=(1/G) | Xn=cosϴ |
| sinϴ | Y=0 | Yn=sinϴ |
| tanϴ | Z=ϴ | tanϴ=( sinϴ/ cosϴ) |
| polar to rectangular | X=R<br>Y=0<br>Z=ϴ | Xn=Rcosϴ<br>Yn=Rsinϴ |

### 3.2.3 Three-D CORDIC

To implement 3-D CORDIC Processor two cascade 2-D CORDIC processors are used; the new processor called sphere and shown in the following figure:

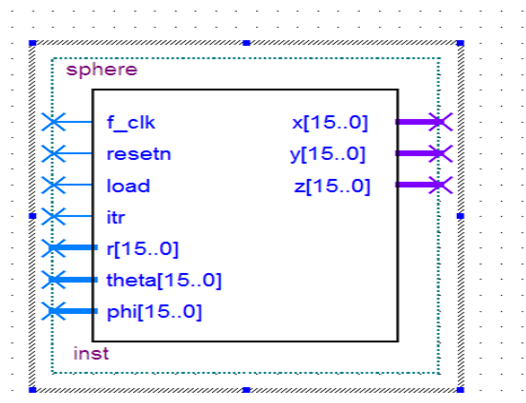

**Fig 7:** 3-D CORDIC Entity using VHDL

Let the spherical coordinates (r, θ, φ) use 3-D CORDIC to get the equivalent Cartesian coordinates(x,y,z) performs in two stages as below :

**The first 2-D CORDIC** has the following inputs:

x0= r ; y0=0 z0= θ  and the output of the first stage will be **rcosθ  and rsinθ** which can be used as inputs to **the second** stage as follows:

x0= rsinθ ; y0=0 , z0=φ and the ouputs will be **x= rsinθcosφ**

**and y=rsinθsinφ** and output x from the first stage represents

**z = rcosθ** .

### 3.3 The Gain Consideration

When the CORDIC algorithm is applied to a vector V to rotate it by an angle Θ, a new vector V´ will be generated in such a way that changes from V to V´ but magnitude change:

Where n is the number of iterations needed to generate the end vector.

This type of change will affect both initial vector components by a constant factor for a fixed number of iterations, let this factor be called the gain Gn, therefore the gain will be given as in eq. 17.

And it will affect both components of V´ by multiplications aggregate to:

Gn = 1.646760258……. as n→ ∞

But CORDIC Rotation mode equations are [6]:

$$x_n = G_n[x_0 \cos z_0 - y_0 \sin z_0] \quad (14)$$

$$y_n = G_n[y_0 \cos z_0 + x_0 \sin z_0] \quad (15)$$

$$z_n = 0 \quad (16)$$

$$Gn = \prod^N \sqrt{1 + 2^{-2i}} \quad (17)$$

Because of implementation of CORDIC twice in the proposed 3-D model, the gain must considered twice to eliminate the gain factor R must set to:

R = (1/Gn) * (1/Gn) = (.607253)^2 = (0.368756206)*10000 ((to eliminate Floating Point)) = 3687.5 = (0E68)$_{hex.}$

### 4. Results and Discussion

The input of 2-D CORDIC is **X0=(1/Gn) and Y0=0 and Z =Θ to get the** desired output  **Xn=cos(Θ) and Yn=sin(Θ)** after 15 iteration .

$$|V'| = \prod^N \sqrt{1 + 2^{-2i}} \, |V| \quad (13) [6]$$

### 4.1. 2-D CORDIC Results:

**Simply let X0= (1/Gn) = (.607253) = (17B9)$_{hexa}$, and Y0=0**, so the above  the final iteration results will be :

$$x_n = \cos z_0$$
$$y_n = x_0 \sin z_0$$
$$z_n = 0$$

**Table 4:** 2-D CORDIC Results when X0=(1/Gn)  and Y0=0 and Z =Θ .

| Z decimal | Z hexadecimal | X Theoretical Cos(z) | X Simulation Cos(z) | X Error *10-4 | Y Theoretical Sine(Z) | Y Simulation Sine(Z) | Y Error *10-4 |
|---|---|---|---|---|---|---|---|
| 90 | 2000 | 0000 | 0002 | 2 | 2710 | 2711 | 1 |
| 75 | 1AAA | 0A1C | 0A1C | 0 | 25BB | 25BD | 2 |
| 60 | 1555 | 1388 | 1387 | 1 | 21D4 | 21D6 | 2 |
| 45 | 1000 | 1BF9 | 1BF9 | 0 | 1BF9 | 1BF9 | 0 |
| 30 | 0AAA | 21D4 | 21D7 | 3 | 1388 | 1386 | 2 |
| 15 | 0555 | 25BB | 25BE | 3 | 0A1C | 0A19 | 3 |
| -15 | FAAA | 25BB | 25BB | 0 | F5E4 | F5E3 | 1 |
| -30 | F555 | 21D4 | 21D6 | 2 | EC78 | EC79 | 1 |
| -45 | F000 | 1B9F | 1BA0 | 1 | E461 | E461 | 0 |
| -60 | EAAA | 1388 | 1387 | 1 | DE2C | DE29 | 3 |
| -75 | E555 | 0A1C | 0A1F | 3 | DA45 | DA42 | 3 |
| -90 | E000 | 0000 | 0000 | 0 | D8F0 | D8EE | 2 |
| Average | | | | 1.33 | | | 1.6667 |

The error in X value (Cos Θ) and error in Y Value (sin Θ) is shown in figure 8 below:

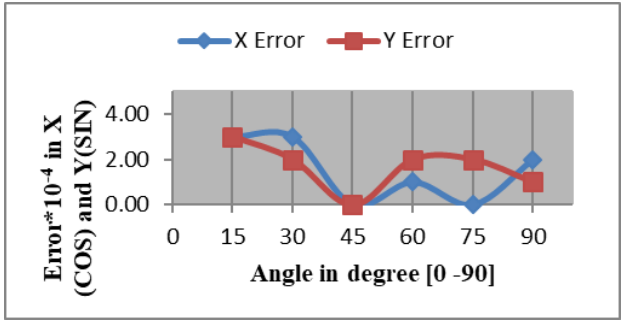

**Fig 8:** The proposed 2-D CORDIC model Error in X_Value (COS (z)) and Y_Value (SIN(Z)).

### 4.2. 3-D CORDIC Results:

Let the spherical coordinates r=0E68, θ=1000, φ=1000 find the equivalent Cartesian coordinates (x, y, z); by calculations:

X_ calculated= r * (sin 45) * (cos 45)= [(1/Gn^2)] * sin45 * cos45=0.5 *10000 =5000=(1388)$_{hexa}$

Y_ calculated = r * (sin 45) * (sin 45)= [(1/Gn^2)] * sin45 * sin45=0.5 *10000 =5000=(1388)$_{hexa}$

Z_ calculated = r * (cos 45) = [(1/Gn^2)] * cos45=0.429455 *10000=4294= (10C6)$_{hexa}$

The simulation result on Quartus II 7.1 software when (r=0E68, θ=1000, φ=1000)as shown in Figure 9:

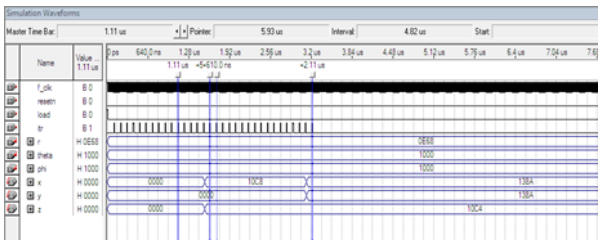

**Fig 9:** Simulation results when(r=0E68, θ=1000, φ=0AAA).

Let the spherical coordinates r=0E68, θ=1555, φ=0AAA find the equivalent Cartesian coordinates (x, y, z); by calculations:

X= r * (sin 60) * (cos 30)= [(1/Gn^2)] * sin60* cos30=0.75 *10000 =7500=(1D4C)$_{hex.}$

Y= r * (sin 60) * (sin 30)= [(1/Gn^2)] * sin60 * sin30=0.433012 *10000 =4330=(10EA)$_{hex.}$

Z= r * (cos 60) = [(1/Gn^2)] * cos60=0.3036707 *10000=3036= (0BDC) $_{hex.}$

The simulation result on Quartus II 7.1 software when (r=0E68, θ=1555, φ=0AAA)as shown in Figure 10:

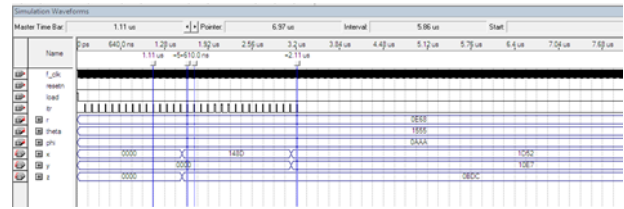

**Figure 10:** Simulation results when(r=0E68, θ=1555, φ=0AAA).

The simulation results for converting spherical coordinate's r=0E68, θ, φ to equivalent Cartesian Coordinates is **summarized in table 5.**

The accuracy of 2-D CORDIC Processor is acceptable; since the average error in cosine(Θ)= $1.33*10^{-4}$ and the average error in sine(Θ) =$1.6667*10^{-4}$.

The accuracy of 3-D CORDIC Processor is acceptable;

Since the average error in X_Value= $4*10^{-4}$

And the average error in Y_Value= $2*10^{-4}$

And the average error in Z_Value= $1*10^{-4}$

**Table 5:** 3-D CORDIC Results Cartesian Coordinates (X, Y, Z) Equivalent to Spherical Coordinates (r=0E68,and θ, φ).

| θ | φ | X Calculated | X simulation | X Deviation*10-4 | Y calculated | Y Simulation | Y De*10-4 | Z Calculate | Z simulation | Z Deviation*10-4 |
|---|---|---|---|---|---|---|---|---|---|---|
| 60 | 30 | 1D4C | 1D52 | **6** | 10EA | 1.00E+08 | **3** | 0BDC | 0BDC | **0** |
| 45 | 30 | 17EC | 17F0 | **4** | 0DCF | 10C5 | **1** | 10C5 | 10C4 | **1** |
| 45 | 45 | 1388 | 138A | **2** | 138A | 10C5 | **2** | 10C6 | 10C4 | **2** |
| Avg | | | | **4** | | | **2** | | | **1** |

The proposed model evaluates high speed calculation for : 2-D CORDIC Processor the latency = 16 clock cycle and for 3-D CORDIC Processor the latency =32 clock cycle .

### 5. Conclusions

This research paper focuses on implementing 3-D CORDIC processor using two 2-D CORDIC Processor. This 3-D CORDIC Processor is used to convert from spherical coordinates to Cartesian coordinates and constructed using VHDL.

Using FPGA as flexible implementation tool make it easy to modify the proposed model without additional cost.

The accuracy of 2-D CORDIC Processor is acceptable; since the average error in cosine($\Theta$)= $1.33*10^{-4}$ and the average error in sine($\Theta$) =$1.6667*10^{-4}$.

The accuracy of 3-D CORDIC Processor is acceptable.
The proposed model evaluates high speed calculation, reducing latency at the cost of accuracy is beneficial for numerous real-time applications like: Digital Signal Processing (DSP), wireless communication systems, navigation and GPS.

**Nomenclatures**

| | |
|---|---|
| $K$ | Constant gain |

**Abbreviations**

| | |
|---|---|
| CORDIC | Coordinate Rotation Digital Computer |
| FPGA | Field Programmable Gate Array |
| LUT | Look Up Table |
| VHDL | Very High-Speed Integrated Circuit Hardware Description Language |